\documentclass[a4paper,11pt]{article}
\pdfoutput=1
\usepackage{jheppub}

\usepackage[utf8]{inputenc}
\usepackage{subfig}
\usepackage{color}
\usepackage{verbatim}
\usepackage{textcomp}
\usepackage{amsmath}
\usepackage{amssymb}
\usepackage{graphicx}
\usepackage{nicefrac}
\usepackage[font=footnotesize]{caption}

\usepackage{array}
\usepackage{multirow}

\usepackage{slashed}
\usepackage{cancel}

\title{Running of fermion observables in non-supersymmetric SO(10) models}

\author[a,b,c]{\textbf{Tommy Ohlsson}}
\author[a,b]{\textbf{and Marcus Pernow}}

\affiliation[a]{Department of Physics, School of Engineering Sciences,\\ KTH Royal Institute of Technology, AlbaNova University Center,\\ Roslagstullsbacken 21, SE-106 91 Stockholm, Sweden}

\affiliation[b]{The Oskar Klein Centre for Cosmoparticle Physics, AlbaNova University Center,\\ Roslagstullsbacken 21, SE-106 91 Stockholm, Sweden}

\affiliation[c]{University of Iceland, Science Institute,\\ Dunhaga 3, IS-107 Reykjavik, Iceland}

\emailAdd{tohlsson@kth.se}
\emailAdd{pernow@kth.se}

\abstract{
We investigate the complete renormalization group running of fermion observables in two different realistic non-su\-per\-sym\-met\-ric models based on the gauge group $\textrm{SO}(10)$ with intermediate symmetry breaking for both normal and inverted neutrino mass orderings. Contrary to results of previous works, we find that the model with the more minimal Yukawa sector of the Lagrangian fails to reproduce the measured values of observables at the electroweak scale, whereas the model with the more extended Yukawa sector can do so if the neutrino masses have normal ordering. The difficulty in finding acceptable fits to measured data is a result of the added complexity from the effect of an intermediate symmetry breaking as well as tension in the value of the leptonic mixing angle $\theta^\ell_{23}$.}


\arxivnumber{1804.04560}

\makeatother

\begin{document}

\maketitle

\section{Introduction}

Grand unified theories provide an intriguing framework for physics beyond the Standard Model (SM). The $\textrm{SO}(10)$ gauge group is a popular version since it accommodates all SM fermions and the right-handed neutrino in one representation~\cite{Georgi:1975qb,Fritzsch:1974nn}. However, in order to be a viable candidate, it must be able to reproduce the experimentally measured fermion masses and mixing parameters. Therefore, it is relevant to analyze how well the parameter values of a particular model can be fitted to the measured observables.

The issue of fermion masses and mixing parameters in non-supersymmetric (non-SUSY) $\textrm{SO}(10)$ frameworks has been extensively discussed previously in the literature, see for example refs.~\cite{Harvey:1981hk,Babu:1992ia,Bajc:2005zf,Witten1980,Matsuda2001,Matsuda2002,Fukuyama2018}. The most minimal choice of scalar representations in the Yukawa sector of the Lagrangian that can reproduce the desired fermion data are the $\mathbf{10}_H$ and $\overline{\mathbf{126}}_H$ representations, which has been demonstrated in a number of previous fits~\cite{Bertolini:2006pe,Joshipura:2011nn,Altarelli:2013aqa,Dueck:2013gca,Meloni:2014rga,Babu:2015bna}. One can also choose to extend the Yukawa sector by adding a $\mathbf{120}_H$ representation~\cite{Joshipura:2011nn,Dueck:2013gca,Babu:2016bmy,Meloni:2016rnt}.

In order to compare the parameters of a high-energy theory to low-energy observables, one must take into account the renormalization group equations (RGEs)~\cite{Ohlsson:2013xva}. Most previous analyses of fermion observables in $\textrm{SO}(10)$ models use solutions of the RGEs for the SM to compare the parameters at the $\textrm{SO}(10)$ breaking scale $M_\textrm{GUT}$ to observables extrapolated from the experimental energy scale up to that scale~\cite{Joshipura:2011nn,Altarelli:2013aqa} or solve the RGEs for the parameter values from $M_\mathrm{GUT}$ down to the electroweak scale $M_\mathrm{Z}$~\cite{Dueck:2013gca}, assuming an SM-like model in the whole energy range. However, non-supersymmetric $\textrm{SO}(10)$ models require an intermediate symmetry breaking~\cite{delAguila:1980qag}, so it is worthwhile to consider a more complete analysis that takes into account the effects of an intermediate gauge group (ignoring it amounts to assuming that its effect is negligible, for example if the associated energy scale $M_\textrm{I}$ of the intermediate symmetry breaking is very close to $M_\textrm{GUT}$). Various breaking chains are possible which have different renormalization group (RG) running of the gauge couplings, resulting in different values for the energy scales $M_\textrm{I}$ and $M_\textrm{GUT}$~\cite{Deshpande:1992au,Bertolini:2009qj}. A commonly considered intermediate symmetry is the Pati--Salam (PS) gauge group~\cite{Pati:1974yy}. The derivation of the complete set of RGEs for the gauge, Yukawa, and scalar couplings~\cite{Jones:1981we,Machacek:1983tz,Machacek:1983fi,Machacek:1984zw} in such a breaking chain as well as their matching conditions at $M_\textrm{I}$ was first attempted in ref.~\cite{Fukuyama:2002vv}. A numerical analysis based on the RGEs and matching conditions presented therein demonstrated the substantial effect that an intermediate gauge group in the symmetry breaking can have on the RG running and fits to fermion observables in a minimal $\textrm{SO}(10)$ model~\cite{Meloni:2014rga}. This analysis was later refined by deriving correct RGEs and also considering an extended (or non-minimal) $\textrm{SO}(10)$ model~\cite{Meloni:2016rnt}. 

The present work aims to extend the analysis of the two models in refs.~\cite{Meloni:2014rga,Meloni:2016rnt} in several ways. Firstly, we fit to the two neutrino mass-squared differences separately, whereas the above mentioned works performed the fits to only their ratio. Secondly, we consider both normal ordering (NO) and inverted ordering (IO) of the neutrino masses. Thus we consider four different cases, namely two different models, each with both NO and IO. Lastly, we update the values of the observables at $M_\textrm{Z}$ to the best-known values to date.

This paper is organized as follows. First, in section~\ref{model}, we briefly describe the models including the breaking chain down to $M_\textrm{Z}$. Then in section~\ref{procedure}, we describe the procedure used to perform the analysis. Next, in section~\ref{results}, the results of the analysis are presented, discussed, and compared to previous results. Finally, in section~\ref{conclusion}, we summarize our findings and conclude.

\section{Description of the minimal and extended models}\label{model}
\begin{figure}
\centering
\includegraphics[width=0.5\textwidth]{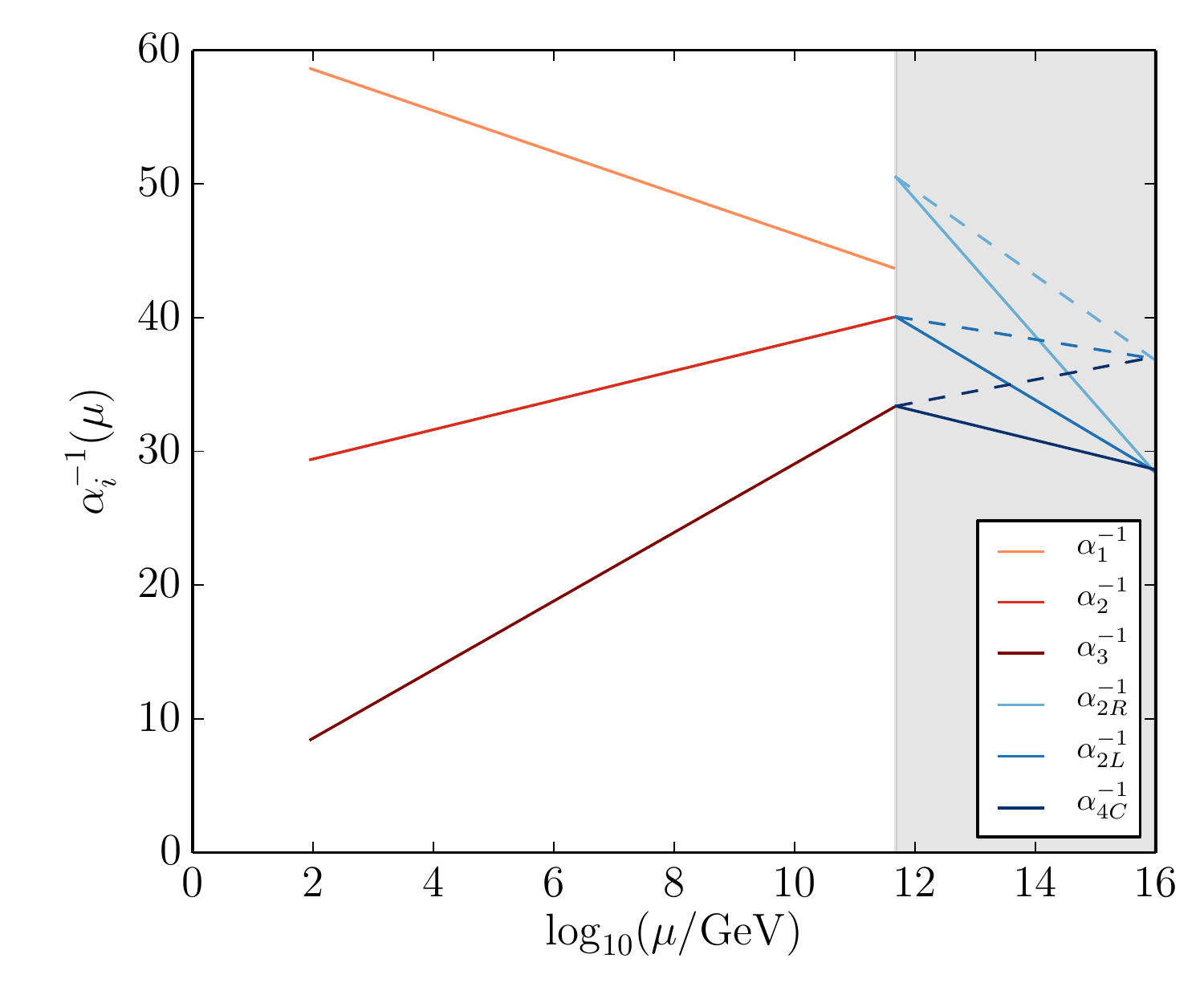}
\caption{\label{fig:gauge} Gauge coupling unification in the minimal model (dashed curves) and extended model (solid curves). Below $M_\mathrm{I}$, the RG running is the same. The shaded region, from $M_\mathrm{I}$ to $M_\mathrm{GUT}$, denotes the energy interval of the intermediate gauge group.}
\end{figure}

In this section, we briefly outline the two models to which fits will be performed. More details on these models can be found in refs.~\cite{Meloni:2014rga, Meloni:2016rnt}. The two models are both non-supersymmetric and based on the $\textrm{SO}(10)$ gauge group. In what follows, they are referred to as the \emph{minimal model} and the \emph{extended model} due to their difference in scalar representations (whether or not the $\mathbf{120}_H$ is included). We assume that the $\textrm{SO}(10)$ symmetry breaking to the spontaneously broken SM in both models proceeds via the PS group, {\it viz.}
\begin{equation}
\begin{split}
\textrm{SO}(10) &\xrightarrow{M_{\textrm{GUT}}} \textrm{SU}(4)_C\otimes\textrm{SU}(2)_L\otimes\textrm{SU}(2)_R\\
&\xrightarrow{\,\,\,\, M_{\textrm{I}}\,\,\,\,} \textrm{SU}(3)_C\otimes\textrm{SU}(2)_L\otimes\textrm{U}(1)_Y\\
&\xrightarrow{\,\,\,\, M_\mathrm{Z}\,\,\,\,} \textrm{SU}(3)_C\otimes\textrm{U}(1)_\mathrm{em}.
\end{split}
\end{equation}
The electroweak symmetry breaking scale is $M_\mathrm{Z}=91.1876 \allowbreak \textrm{GeV}$~\cite{Patrignani:2016xqp} and the energy scales of the other two symmetry breakings are computed to be~\cite{Meloni:2016rnt}
\begin{equation}
M_{\textrm{I}}=4.8\cdot 10^{11} \, \textrm{GeV}\quad \mbox{and} \quad M_{\textrm{GUT}}=10^{16} \, \textrm{GeV},
\end{equation}
respectively.
These energy scales are uniquely derived from the requirement of gauge coupling unification at $M_{\rm{GUT}}$ with the coupling constants  
\begin{equation}
\alpha_i^{-1}(M_{\rm{GUT}})\simeq
\begin{cases}
37.0 \quad \textrm{(minimal model)} \\
28.6 \quad \textrm{(extended model)}
\end{cases},
\end{equation}
as shown in figure~\ref{fig:gauge}, where the convention $\alpha_i=g_i^2/(4\pi)$ has been used. Note that we can perform this analysis independent of the RG running of the Yukawa couplings since, to one-loop order, the RGEs for the gauge couplings are independent of those of the Yukawa couplings~\cite{Jones:1981we,Machacek:1983tz}.

\subsection{SO(10) Lagrangians}
Above $M_{\textrm{GUT}}$, the Yukawa sector of the Lagrangian for the minimal model is given by
\begin{equation}
-\mathcal{L}_Y^{\rm{GUT, min}}=\mathbf{16}_F(h\mathbf{10}_H+f\mathbf{\overline{126}}_H)\mathbf{16}_F,
\end{equation}
where $\mathbf{16}_F$ is the spinor representation containing the fer\-mions, whereas $\mathbf{10}_H$ and $\mathbf{\overline{126}}_H$ contain the Higgs scalars. Note that we forbid the coupling to the conjugate $\mathbf{10}^*_H$ by imposing a Peccei-Quinn $\textrm{U}(1)_\textrm{PQ}$ symmetry~\cite{Babu:1992ia,Bajc:2005zf}. In the extended model, we also include the  $\mathbf{120}_H$ Higgs representation. Therefore, the Yukawa sector of the Lagrangian for this model is 
\begin{equation}
-\mathcal{L}_Y^{\rm{GUT, ext}}=\mathbf{16}_F(h\mathbf{10}_H+f\overline{\mathbf{126}}_H+g\mathbf{120}_H)\mathbf{16}_F.
\end{equation}
The Yukawa couplings $h$, $f$, and $g$ are $3\times 3$ matrices in flavor space. For simplicity, one can choose a basis in which $h$ is real and diagonal. The other two matrices $f$ and $g$ are then complex symmetric and complex antisymmetric, respectively. 

\subsection{Pati-Salam Lagrangians}
Between $M_{\textrm{GUT}}$ and $M_{\textrm{I}}$, the relevant fields decompose in the PS group as
\begin{equation}
\begin{split}
\mathbf{16}_F&=(\mathbf{4},\mathbf{2},\mathbf{1}) \oplus (\overline{\mathbf{4}},\mathbf{1},\mathbf{2})\equiv F_L \oplus F_R,\\
\mathbf{10}_H&=(\mathbf{1},\mathbf{2},\mathbf{2}) \oplus (\mathbf{6},\mathbf{1},\mathbf{1}),\\
\overline{\mathbf{126}}_H&=(\mathbf{6},\mathbf{1},\mathbf{1}) \oplus (\mathbf{10},\mathbf{1},\mathbf{3}) \oplus (\overline{\mathbf{10}},\mathbf{3},\mathbf{1}) \oplus (\mathbf{15},\mathbf{2},\mathbf{2}),\\
\mathbf{120}_H&=(\mathbf{10}+\overline{\mathbf{10}},\mathbf{1},\mathbf{1}) \oplus (\mathbf{6},\mathbf{3},\mathbf{1}) \oplus (\mathbf{6},\mathbf{1},\mathbf{3}) \oplus (\mathbf{15},\mathbf{2},\mathbf{2}) \oplus (\mathbf{1},\mathbf{2},\mathbf{2}).
\end{split}
\end{equation}
The fields that contribute to the particle masses are 
\begin{equation}
\begin{split}
&\Phi_{10}\equiv (\mathbf{1},\mathbf{2},\mathbf{2})_{10},\quad \Sigma_{126} \equiv (\mathbf{15},\mathbf{2},\mathbf{2})_{126}, \\
&\Phi_{120}\equiv (\mathbf{1},\mathbf{2},\mathbf{2})_{120},\quad \Sigma_{120}\equiv (\mathbf{15},\mathbf{2},\mathbf{2})_{120}, \\
&\overline{\Delta}_R\equiv (\mathbf{10},\mathbf{1},\mathbf{3})_{126},
\end{split}
\end{equation}
where the subscripts indicate which representation they originate from. Since these are the only scalars involved in the breaking chains, we appeal to the extended survival hypothesis to assume that they are the only ones that are present at this scale~\cite{delAguila:1980qag,Dimopoulos:1984ha,Altarelli:2013aqa}. The Lagrangian for the minimal model between $M_\mathrm{GUT}$ and $M_\mathrm{I}$ is chosen as
\begin{equation}
-\mathcal{L}_Y^{\textrm{PS, min}}=Y_F^{(10)}\overline{F}_L\Phi_{10}F_R + Y_F^{(126)}\overline{F}_L\Sigma_{126}F_R+Y_R^{(126)}F_R^TCF_R\overline{\Delta}_R,
\end{equation}
whereas for the extended model, we choose
\begin{equation}
\begin{split}
-\mathcal{L}_Y^{\textrm{PS, ext}}&=Y_F^{(10)}\overline{F}_L\Phi_{10}F_R + Y_F^{(126)}\overline{F}_L\Sigma_{126}F_R\\
&+Y_R^{(126)}F_R^TCF_R\overline{\Delta}_R+Y_{F,1}^{(120)}\overline{F}_L\Phi_{120}F_R\\
&+Y_{F,2}^{(120)}\overline{F}_L\Sigma_{120}F_R,
\end{split}
\end{equation}
where $C$ is the charge-conjugation matrix. The Yukawa coupling matrices $Y_F^{(10)}$, $Y_F^{(126)}$, $Y_R^{(126)}$, $Y_{F,1}^{(120)}$, and $Y_{F,2}^{(120)}$ are related to the ones appearing in the $\textrm{SO}(10)$ Lagrangians by a set of matching conditions~\cite{Hall:1980kf,Deshpande:1992au,Fukuyama:2002vv}, for which we refer the reader to refs.~\cite{Meloni:2014rga,Meloni:2016rnt}. In ref.~\cite{Meloni:2016rnt}, the correct RGEs can also be found, which determine the evolution of the gauge and Yukawa couplings between $M_{\textrm{GUT}}$ and $M_{\textrm{I}}$.

\subsection{SM-like Lagrangian}\label{2HDM}
Below $M_{\textrm{I}}$ (and above $M_{\textrm{Z}}$), we assume as an SM-like model a two-Higgs-doublet model (2HDM), since this is embedded naturally into the PS model. The Lagrangian of the Yukawa sector is thus 
\begin{equation}
-\mathcal{L}_Y^{\textrm{2HDM}}=Y_u\overline{q}_L\phi_2u_R + Y_d\overline{q}_L\phi_1d_R + Y_e\overline{\ell}_L\phi_1e_R + Y_D\overline{\ell}_L\phi_2N_R
\end{equation}
for both the minimal and extended models. Here, $q_L$ and $\ell_L$ are the quark and lepton $\textrm{SU}(2)_L$ doublets, respectively, and $u_R$, $d_R$, $e_R$, and $N_R$ are the quark and lepton $\textrm{SU}(2)_L$ singlets, respectively. The coefficients $Y_u$, $Y_d$, $Y_e$, and $Y_D$ are Yukawa matrices for the up-type quarks, down-type quarks, charged leptons, and neutrinos, respectively, and $\phi_1$ and $\phi_2$ are the two Higgs scalars. 
The vacuum expectation values (vevs), which are involved in the matching conditions for the Yuk\-awa matrices~\cite{Meloni:2016rnt}, are denoted as
\begin{equation}
\begin{split}
&k_{u,d}=\left\langle \Phi_{10} \right\rangle_{u,d}, \; v_{u,d}=\left\langle \Sigma_{126} \right\rangle_{u,d}, \; v_R=\left\langle \overline{\Delta}_R \right\rangle,\\
&z_{u,d}=\left\langle \Phi_{120} \right\rangle_{u,d}, \; t_{u,d}=\left\langle \Sigma_{120} \right\rangle_{u,d},
\end{split}
\label{eq:vevs}
\end{equation}
where $\overline{\Delta}_R$ takes vev at $M_\mathrm{I}$, while the others are $\mathrm{SU}(2)_L$ doublets which take vev at $M_\mathrm{Z}$. Note that similarly to ref.~\cite{Meloni:2016rnt}, we make the simplifying assumption that, although all $\mathrm{SU}(2)_L$ doublet scalars contribute to the fermion masses, the Higgs doublets $\phi_1$ and $\phi_2$ consist predominantly of the two doublet scalars originating in $\Phi_{10}$. The constraint on the vevs can therefore be approximated to $\sqrt{k_u^2+k_d^2}=246\,\textrm{GeV}$~\cite{Branco:2011iw}. A more complete analysis of the scalar potential is needed to determine the composition of $\phi_1$ and $\phi_2$ in terms of the available scalars and may result in a different constraint on the vevs. However, this is beyond the scope of our work and we make the assumption that the dominant contributions to $\phi_1$ and $\phi_2$ come from $\Phi_{10}$.

The RGEs for the evolution of the gauge and Yukawa couplings have previously been presented in the literature~\cite{Fukuyama:2002vv,Meloni:2016rnt}. For neutrino masses, we assume a type-I seesaw mechanism with the seesaw scale close to $M_{\textrm{I}}$. Thus, we have an effective neutrino mass matrix 
\begin{equation}
m_{\nu}=M_D^TM_R^{-1}M_D,
\end{equation}
where $M_D=(k_u/\sqrt{2})Y_D$ is the Dirac neutrino mass matrix and $M_R$ is the right-handed Majorana neutrino mass matrix. For more details on its relation to the Yukawa couplings in the PS model as well as details regarding the RG running of neutrino parameters, the reader is referred to ref.~\cite{Meloni:2016rnt}. As explained therein, we also need to include a Higgs self-coupling for each Higgs doublet, since they affect the RG running of the neutrino mass matrix~\cite{Antusch:2001vn,Grimus:2004yh}. However, as in ref.~\cite{Meloni:2016rnt}, we assume that the quartic couplings that involve cross-couplings between the two Higgs doublets are zero, so that we have the scalar potential $V(\phi_1,\phi_2)=\lambda_1 (\phi_1^\dagger\phi_1)^2+\lambda_2 (\phi_2^\dagger\phi_2)^2$.

A summary of the relevant quantities in the different energy regions is presented in table~\ref{tab:quantities}. At each symmetry breaking scale, the Yukawa couplings of the lower energy theory are linear combinations of those of the higher energy theory. The quantities that exhibit RG running --- \emph{and therefore change with energy} --- in the PS model are the Yukawa couplings in $M_\mathrm{I} < \mu < M_\mathrm{GUT}$ as well as the gauge couplings. In the 2HDM, the quantities that exhibit RG running --- \emph{and therefore change with energy} --- are the Yukawa couplings in $M_\mathrm{Z} < \mu < M_\mathrm{I}$ as well as the effective neutrino mass matrix, gauge couplings, and Higgs quartic couplings. The only vevs that we consider are those that contribute to the fermion masses, which are listed in eq.~\eqref{eq:vevs}. All except $v_R$ are vevs of $\mathrm{SU}(2)_L$ doublet scalars which take vevs at $M_\mathrm{Z}$, but enter as effective parameters in the matching of Yukawa couplings at $M_\mathrm{I}$. We assume that the vevs are constant in energy.

\begin{table}[t]
\centering
\begin{tabular}{|l l l|}
\hline 

Energy region 															& Yukawa couplings 		& Scalar fields								\tabularnewline
\hline 
$M_\mathrm{GUT} < \mu$ 		 				  			& $h$, $f$, $g$			 	& $\mathbf{10}_H$, $\mathbf{\overline{126}}_H$, $\mathbf{120}_H$		\tabularnewline
$M_\mathrm{I} < \mu < M_\mathrm{GUT}$			& \vtop{\hbox{\strut $Y_F^{(10)}$, $Y_F^{(126)}$, $Y_R^{(126)}$}\hbox{\strut $Y_{F,1}^{(120)}$, $Y_{F,2}^{(120)}$}}, \newline 								& \vtop{\hbox{\strut $\Phi_{10}$	, $\Sigma_{126}$, $\Phi_{120}$,}\hbox{\strut $\Sigma_{120}$,	$\overline{\Delta}_R$}} 					\tabularnewline
$M_\mathrm{Z} < \mu < M_\mathrm{I}$ 			  	& $Y_u$, $Y_d$, $Y_e$, $Y_D$ 							& $\phi_1$, $\phi_2$							\tabularnewline
\hline 

\end{tabular}
\caption{\label{tab:quantities}Summary of relevant Yukawa couplings and scalar fields in each energy region.}
\end{table}

\section{Parameter-fitting procedure}\label{procedure}
In this section, we describe the procedure and numerical tools used to perform the parameter fits, which follows closely refs.~\cite{Meloni:2014rga,Meloni:2016rnt}. The general procedure consists of minimizing a $\chi^2$ function, which is formed by comparing measured data at $M_\textrm{Z}$ with the RG running of parameter values from $M_\textrm{GUT}$ to $M_\textrm{Z}$ in a given $\textrm{SO}(10)$ model. This RG running is performed by solving the relevant RGEs of the model parameters from $M_\mathrm{GUT}$ to $M_\mathrm{Z}$, taking into account the change of parameters at $M_\mathrm{I}$. Due to the nature of the matching conditions at $M_\mathrm{I}$, it is not possible to extrapolate the observables from $M_\mathrm{Z}$ to $M_\mathrm{GUT}$ and we are forced to perform the RG running from the high-energy model down to the low-energy observables. Note that due to the intermediate PS symmetry, the parameters for which the RGEs are solved above $M_I$ are not the Yukawa couplings that appear in the SM, but rather the Yukawa couplings $Y_F^{(10)}$, $Y_F^{(126)}$, $Y_R^{(126)}$, $Y_{F,1}^{(120)}$, and $Y_{F,2}^{(120)}$ of the PS model. In the minimal model, there are 22 parameters: three in $h$, twelve in $f$, four in the complex vevs $v_u$ and $v_d$, one in the ratio of the real vevs $k_u/k_d$, one in the real vev $v_R$, and one in the Higgs self-coupling $\lambda$ (since the two are assumed to be equal above $M_{\textrm{I}}$). The extended model has a total of 34 parameters, which are the 22 of the minimal model and an extra twelve: six in $g$, four in the complex vevs $t_u$ and $z_u$, and two in the real vevs $t_d$ and $z_d$.

In order to determine the values of the above-mentioned parameters that provide the best fit to measured data, we employ the following strategy:
\begin{enumerate}
\item Generate the parameter values at $M_\mathrm{GUT}$.
\item Numerically solve the one-loop RGEs of the parameters that exhibit RG running to relate the parameter values at $M_\mathrm{GUT}$ to those at $M_\mathrm{Z}$. At $M_\textrm{I}$, use the matching conditions to transform the parameters to the ones that are relevant in the lower energy region.
\item Construct the 18 fermion observables (masses and mixing parameters) at $M_\textrm{Z}$ and compare these to measured data by calculating the corresponding value of the $\chi^2$ function.
\item Repeat the above steps to find the parameter values the provide the best fit and the corresponding value of the $\chi^2$ function.
\end{enumerate}
The $\chi^2$ function is defined as
\begin{equation}
\chi^2=\sum_{i=1}^N\left(\frac{\mu_i-X_i}{\sigma_i} \right)^2\equiv\sum_{i=1}^Np_i^2,
\label{eq:chi2}
\end{equation}
where $X_i$ is the measured value of the $i$th observable at $M_\mathrm{Z}$ with corresponding error $\sigma_i$ and $\mu_i$ is the corresponding predicted value of the given model for the current choice of parameter values. We also define the pulls $p_i$ as above for later convenience. For the sampling of the parameters, we interchangeably use the packages MultiNest~\cite{Feroz:2007kg,Feroz:2008xx,Feroz:2013hea}, which is a nested sampling algorithm, and Diver~\cite{Workgroup:2017htr}, which is a differential evolution algorithm. Prior distributions are used to generate the next iteration of parameter values such that the elements of $h$, $f$, and $g$ are sampled from logarithmic priors between $10^{-20}$ and $10^{-1}$ (and allowed to be negative), $\lambda$ is sampled from a uniform prior between $-1$ and $1$, and the vevs are sampled from uniform priors between $-550~\textrm{GeV}$ and $550~\textrm{GeV}$, except for $v_R$ which is sampled from a uniform prior between $10^{12}~\textrm{GeV}$ and $10^{16}~\textrm{GeV}$ (this departure from the extended survival hypothesis is necessary to reproduce the neutrino mass-squared differences). The ratio $k_u/k_d$ which is sampled from a uniform prior between $-550$ and $550$. The ranges of the above-mentioned priors are obtained from their expected orders of magnitude as well as preliminary numerical tests. After the sampling algorithm has converged on a set of parameter values, a Nelder-Mead simplex algorithm~\cite{Press:1992zz} is used to further evolve the parameter values to a set that provides an even better fit. However, note that one can never be sure that the global minimum is found. The best that one can do is to restart the minimization procedure several times with different starting parameter values.

\begin{table}[t]
\centering
\begin{tabular}{|l l l r|}
\hline 
Observable 														& $X_i$ 							& $\sigma_i$					&$\sigma_i/X_i$\tabularnewline
\hline 
$m_d$ [GeV] 						  				  			& $2.71\cdot 10^{-3}$ 	& $1.4 \cdot 10^{-3}$		& $50~\%$\tabularnewline
$m_s$ [GeV] 						  				  			& $0.0553$ 					& $0.017$							& $30~\%$\tabularnewline
$m_b$ [GeV] 						  				  			& $2.86$ 						& $0.086 $							& $3~\%$\tabularnewline
$m_u$ [GeV] 						  				  			& $1.27\cdot 10^{-3}$ 	& $6.4\cdot 10^{-4}$			& $50~\%$\tabularnewline
$m_c$ [GeV] 						  				  			& $0.634$					 	& $0.10$								& $15~\%$\tabularnewline
$m_t$ [GeV] 						  				  			& $171$							& $3.5$						 		& $2~\%$\tabularnewline
$\sin\theta_{12}^q$ 			  				  			& $0.225$ 						& $2.3\cdot 10^{-3}$			& $1~\%$\tabularnewline
$\sin\theta_{13}^q$			  				  			& $3.57\cdot 10^{-3}$ 	& $3.6 \cdot 10^{-4}$		& $10~\%$\tabularnewline
$\sin\theta_{23}^q$ 			 				  			& $0.0411$ 					& $1.3\cdot 10^{-3}$			& $3~\%$\tabularnewline
$\delta_\textrm{CKM}$ 		  				  			& $1.24$ 						& $0.062$							& $5~\%$\tabularnewline
$m_e$ [GeV] 						  				  			& $4.87\cdot 10^{-4}$ 	& $2.4 \cdot 10^{-5}$		& $5~\%$\tabularnewline
$m_\mu$ [GeV] 					  				  			& $0.103$ 						& $5.1 \cdot 10^{-3}$		& $5~\%$\tabularnewline
$m_\tau$ [GeV] 					  				  			& $1.75$ 						& $0.087 $							& $5~\%$\tabularnewline
$\Delta m_{21}^2\,[\textrm{eV}^2]$ 			& $7.40\cdot 10^{-5}$ 	& $6.7\cdot 10^{-6}$			& $9~\%$\tabularnewline
$\Delta m_{31}^2\,[\textrm{eV}^2]$ (NO)  & $2.49\cdot 10^{-3}$ 	& $7.5\cdot 10^{-5}$			& $3~\%$\tabularnewline
$\Delta m_{32}^2\,[\textrm{eV}^2]$ (IO)   & $-2.47\cdot 10^{-3}$ & $7.4\cdot 10^{-5}$		& $3~\%$\tabularnewline
$\sin^2\theta_{12}^\ell$ 		 				  			& $0.307$ 					& $0.013$				& $4~\%$\tabularnewline
$\sin^2\theta_{13}^\ell$ (NO)				  			& $0.0221$ 	& $7.5\cdot 10^{-4}$				& $3~\%$\tabularnewline
$\sin^2\theta_{13}^\ell$ (IO) 				 		 	& $0.0223$ 	& $7.4\cdot 10^{-4}$				& $3~\%$\tabularnewline
$\sin^2\theta_{23}^\ell$ (NO)				  			& $0.538$				 	& $0.069$			& $13~\%$\tabularnewline
$\sin^2\theta_{23}^\ell$ (IO) 				  			& $0.554$				 	& $0.033$			& $6~\%$\tabularnewline
\hline 
\end{tabular}
\caption{\label{tab:experimental}Mean values of the 18 observables and corresponding errors at the electroweak scale $M_\textrm{Z}$. The label NO (IO) denotes the parameter values of normal (inverted) neutrino mass ordering. Mean values of the quark and charged-lepton masses are based on updated calculations of refs.~\cite{Xing:2007fb,Xing:2011aa} and mean values of the quark mixing parameters are computed from values given in ref.~\cite{Patrignani:2016xqp}. The neutrino mass-squared differences and the leptonic mixing angles are taken from refs.~\cite{Esteban:2016qun,nu-fit18}.}
\end{table}

In table~\ref{tab:experimental}, we list the measured values of the 18 observables that we fit to. Some comments regarding the choice of values and their corresponding errors are in order. Firstly, the values of the quark and charged-lepton masses are taken from an updated RG running analysis, using the same method as in refs.~\cite{Xing:2007fb,Xing:2011aa}. The relative errors of the quark masses are set to values between $50~\%$ (up and down quarks) and $2~\%$ (top quark), motivated by large theoretical uncertainties in the quark masses, whereas the charged-lepton masses have relative errors set to $5~\%$, due to their almost negligible experimental errors, since otherwise their small errors render the fit practically impossible. Secondly, the values of the quark mixing angles are calculated from the elements of the Cabibbo-Kobayashi-Maskawa (CKM) matrix given in ref.~\cite{Patrignani:2016xqp}, whereas the Dirac CP-violating phase of the CKM matrix is computed from the Wolfenstein parameters of the same reference. The chosen relative errors between $1~\%$ and $10~\%$ reflect the relation among the uncertainties of the quark mixing parameters. Finally, the values of the leptonic mixing angles and the neutrino mass-squared differences are taken from refs.~\cite{Esteban:2016qun,nu-fit18}, as are the associated errors of the leptonic mixing angles. For the mass-squared differences, we choose the relative errors so that their ratio has a relative error of $10~\%$, since the neutrino mass-squared differences have larger uncertainties than the charged-lepton masses. Note that we do not fit to the leptonic Dirac CP-violating phase, since knowledge of its value is limited to indications from global fits, see for example refs.~\cite{Esteban:2016qun,nu-fit18}. 

\begin{table}[t]
\centering
\begin{tabular}{|c c c|}
\hline
Neutrino mass ordering		&		Minimal model	&		Extended model \\
\hline
NO	&		85.9					&		18.6				\\
IO 	&		1424					& 		3081\\
\hline

\end{tabular}
\caption{\label{tab:chi2}Values of the $\chi^2$ function corresponding to the best fits for the two models and for both normal (NO) and inverted (IO) neutrino mass ordering.}
\end{table}

\section{Results and discussion}\label{results}
The $\chi^2$ minimization procedure resulted in only one of the four cases having an acceptable fit, namely the extended model with NO, as shown in table~\ref{tab:chi2}. For this case, the numerical values of the matrices $h$, $f$, and $g$ corresponding to the minimum value of the $\chi^2$ function~are

\begin{equation}
{\small
\begin{split}
h &= \begin{pmatrix}-1.37\cdot10^{-6} & 0 & 0\\
							0 & -1.31\cdot10^{-3} & 0\\
							0 & 0 & 0.300
							\end{pmatrix}, \\
f &= \begin{pmatrix}6.41\cdot10^{-19}-1.32\cdot10^{-25}{\rm i} & 6.60\cdot10^{-6}+1.17\cdot10^{-15}{\rm i} &
 							-4.82\cdot10^{-5}-4.77\cdot10^{-14}{\rm i}\\
							6.60\cdot10^{-6}+1.17\cdot10^{-15}{\rm i} & -1.05\cdot10^{-4}-6.36\cdot10^{-5}{\rm i} &
							 1.33\cdot10^{-12}-1.98\cdot10^{-3}{\rm i}\\
							-4.82\cdot10^{-5}-4.77\cdot10^{-14}{\rm i} & 1.33\cdot10^{-12}-1.98\cdot10^{-3}{\rm i} &
							 -4.16\cdot10^{-12}-7.15\cdot10^{-4}{\rm i}
							\end{pmatrix}, \\
g &= \begin{pmatrix}0 & -1.58\cdot10^{-7}-1.22\cdot10^{-6}{\rm i} &
							 1.40\cdot10^{-14}+1.18\cdot10^{-17}{\rm i}\\
							1.58\cdot10^{-7}+1.22\cdot10^{-6}{\rm i} & 0 & 3.90\cdot10^{-3}+3.79\cdot10^{-13}{\rm i}\\
							-1.40\cdot10^{-14}-1.18\cdot10^{-17}{\rm i} & -3.90\cdot10^{-3}-3.79\cdot10^{-13}{\rm i} & 0
							\end{pmatrix},
\end{split}
}
\label{eq:fgh}
\end{equation}
and the values of the remaining parameters are given in table~\ref{tab:bestfit}.

\begin{table}[h!]
\centering
\begin{tabular}{|l c|}
\hline

Parameter 		& Best-fit value	\\
\hline
$k_u/k_d$				&		$65.5$				\\
$v_u$ [GeV]	&		$0.0 + 4.91\rm{i} $					\\
$v_d$ [GeV]	&		$0.0 -110\rm{i}$				\\
$t_u$ [GeV]	&		$-666 + 0.0\rm{i}$				\\
$t_d$ [GeV]	&		$-149$					\\
$z_u$ [GeV]	&		$ 0.0 + 105 \rm{i}$					\\
$z_d$ [GeV]	&		$56.6$					\\
$\lambda$		&		$ 0.0$					\\
$v_R$	[GeV]	&		$  5.51\cdot 10^{13}$					\\
\hline

\end{tabular}
\caption{\label{tab:bestfit}Parameter values of the extended model with normal neutrino mass ordering corresponding to the minimum value of the $\chi^2$ function.}
\end{table}

\begin{figure}[h]
\centering
\includegraphics[width=0.5\textwidth]{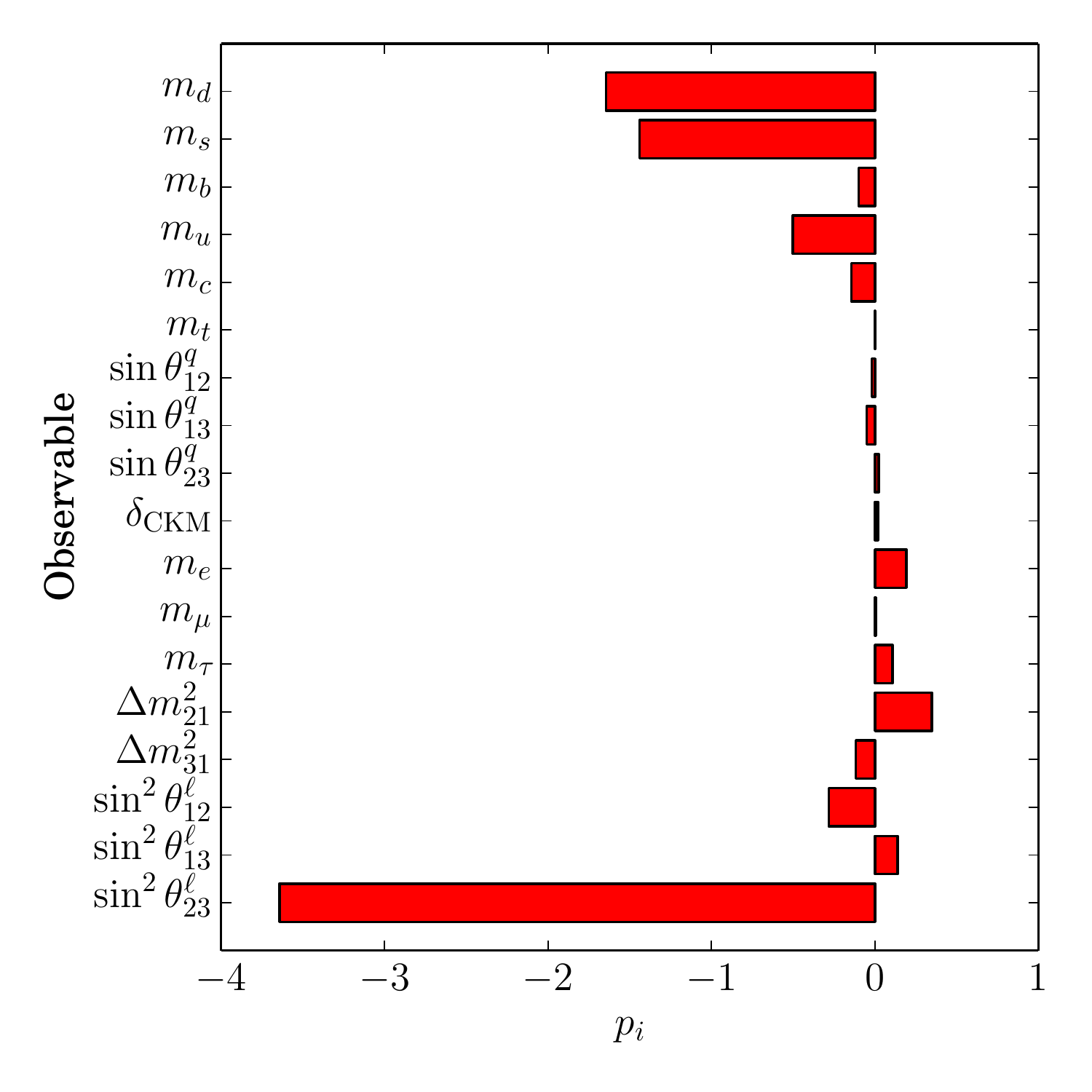}
\caption{\label{fig:pulls}Pulls $p_i$ for each observable in the extended model with normal neutrino mass ordering, given the set of parameter values that has the minimum value of the $\chi^2$ function.}
\end{figure}

In figure~\ref{fig:pulls}, the pulls $p_i$ for each observable, defined in eq.~\eqref{eq:chi2}, are displayed. The sum of the squares of the pulls is the $\chi^2$ function. It is evident that the largest contribution to the $\chi^2$ function is due to the observable $\sin^2\theta^\ell_{23}$, for which the obtained prediction from the best-fit parameters is $0.287$ (corresponding to $\theta^\ell_{23}\simeq32.4^\circ$ in the lower octant), which is significantly lower than the measured value of $0.538$ (corresponding $\theta^\ell_{23}\simeq47.2^\circ$ in the higher octant). This tension due to the octant of $\theta^\ell_{23}$ was also noted in a previous fit to observables at $M_\mathrm{GUT}$~\cite{Altarelli:2013aqa}. In fact, the measured value of $\sin^2\theta^\ell_{23}$ used in the minimization procedure comes from a global fit~\cite{Esteban:2016qun,nu-fit18} and although the $1\sigma$ range does not include our predicted value, it does allow for $\theta^\ell_{23}$ in the lower octant. Furthermore, previous versions of the global fit~\cite{Esteban:2016qun,nu-fit16} predicted a value of $\theta^\ell_{23}$ in the lower octant, which was used in a fit similar to ours presented in ref.~\cite{Meloni:2016rnt}. Replacing the present measured value of $\sin^2\theta^\ell_{23}$ by its previous value of $0.441$, the $\chi^2$ function for our current best-fit parameters takes the value $10.4$, which is lower than the value $11.2$ presented in ref.~\cite{Meloni:2016rnt}. Furthermore, we agree with their conclusion that significant tension in the fit is caused by the masses of the down and strange quarks.

Since the fit to the minimal model with NO is not totally unacceptable, it is worth to consider the significant contributions to its $\chi^2$ function. The largest contribution comes from $\sin^2\theta^\ell_{12}$, followed by $\sin^2\theta^\ell_{23}$. In absolute terms, the best-fit value of $\sin^2\theta^\ell_{12}$ is closer to the measured value than what is the case for $\sin^2\theta^\ell_{23}$, but since the relative error of the former is much smaller than that of the latter, it gives a larger contribution to the value of its $\chi^2$ function. Similarly to the extended model, the best-fit parameter values of this model also predict a value of $\theta^\ell_{23}$ in the lower octant.

\begin{figure*}[t]
\begin{tabular}{c c}
\includegraphics[width=0.49\textwidth]{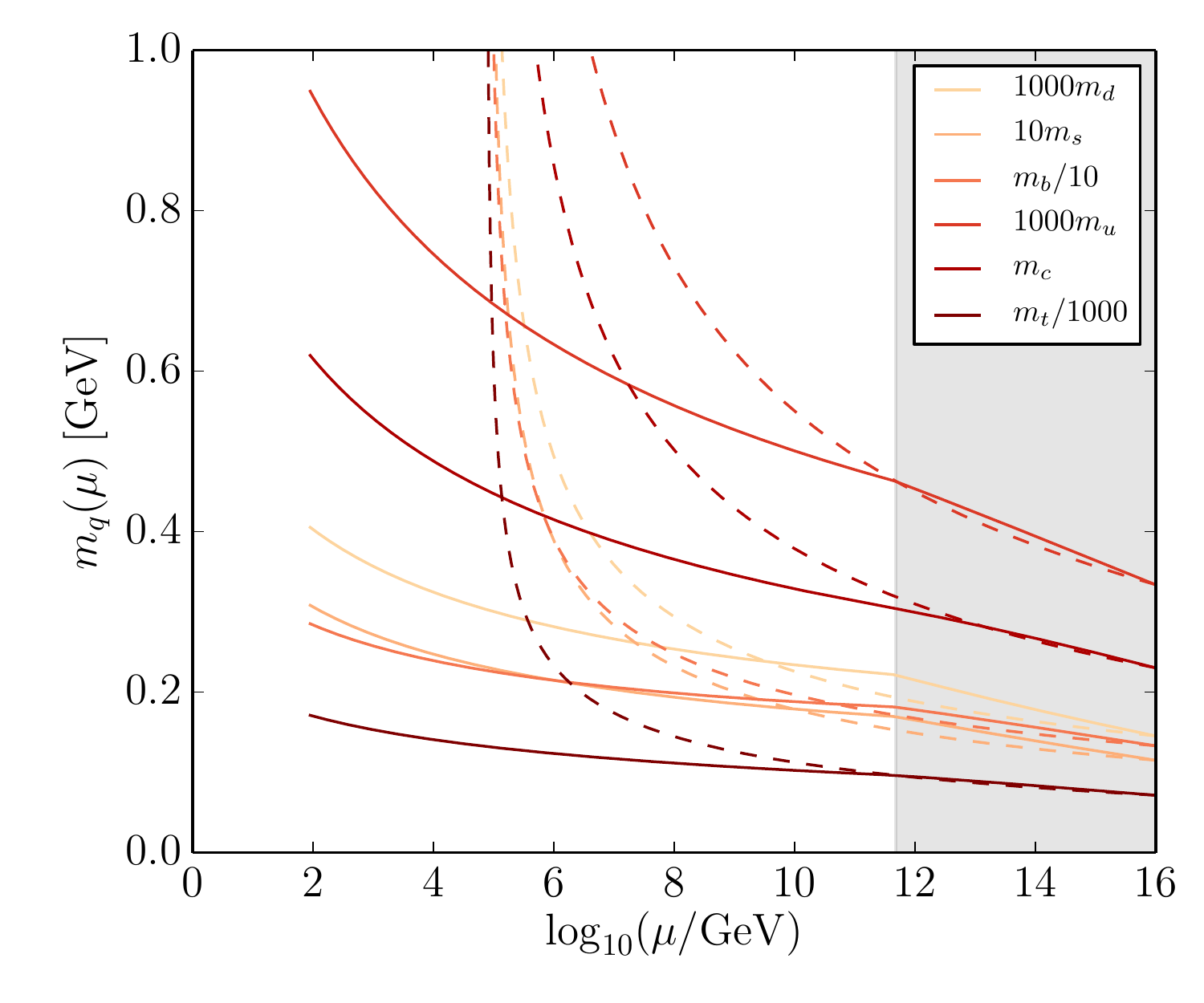} &

\includegraphics[width=0.49\textwidth]{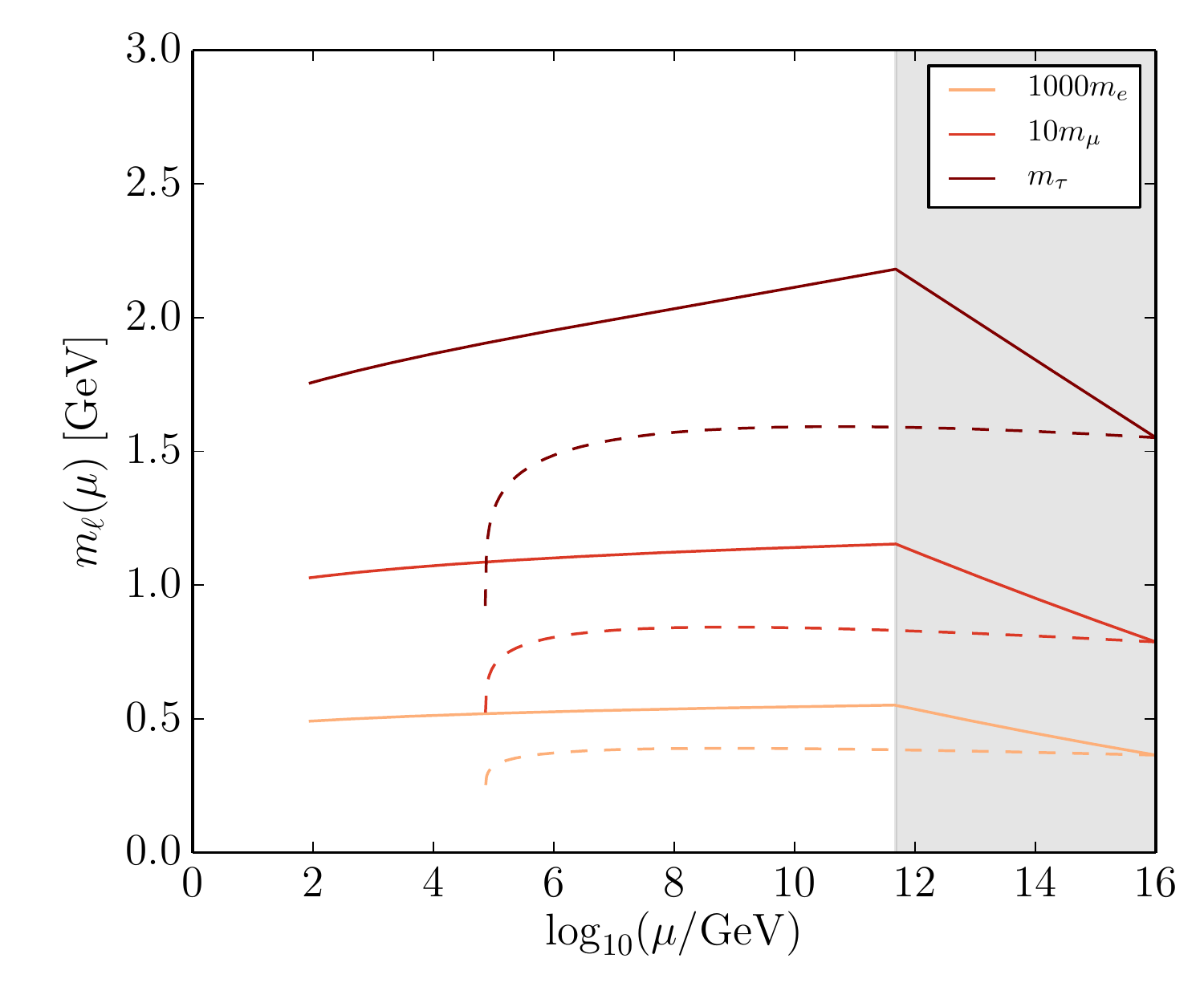}\\

\includegraphics[width=0.49\textwidth]{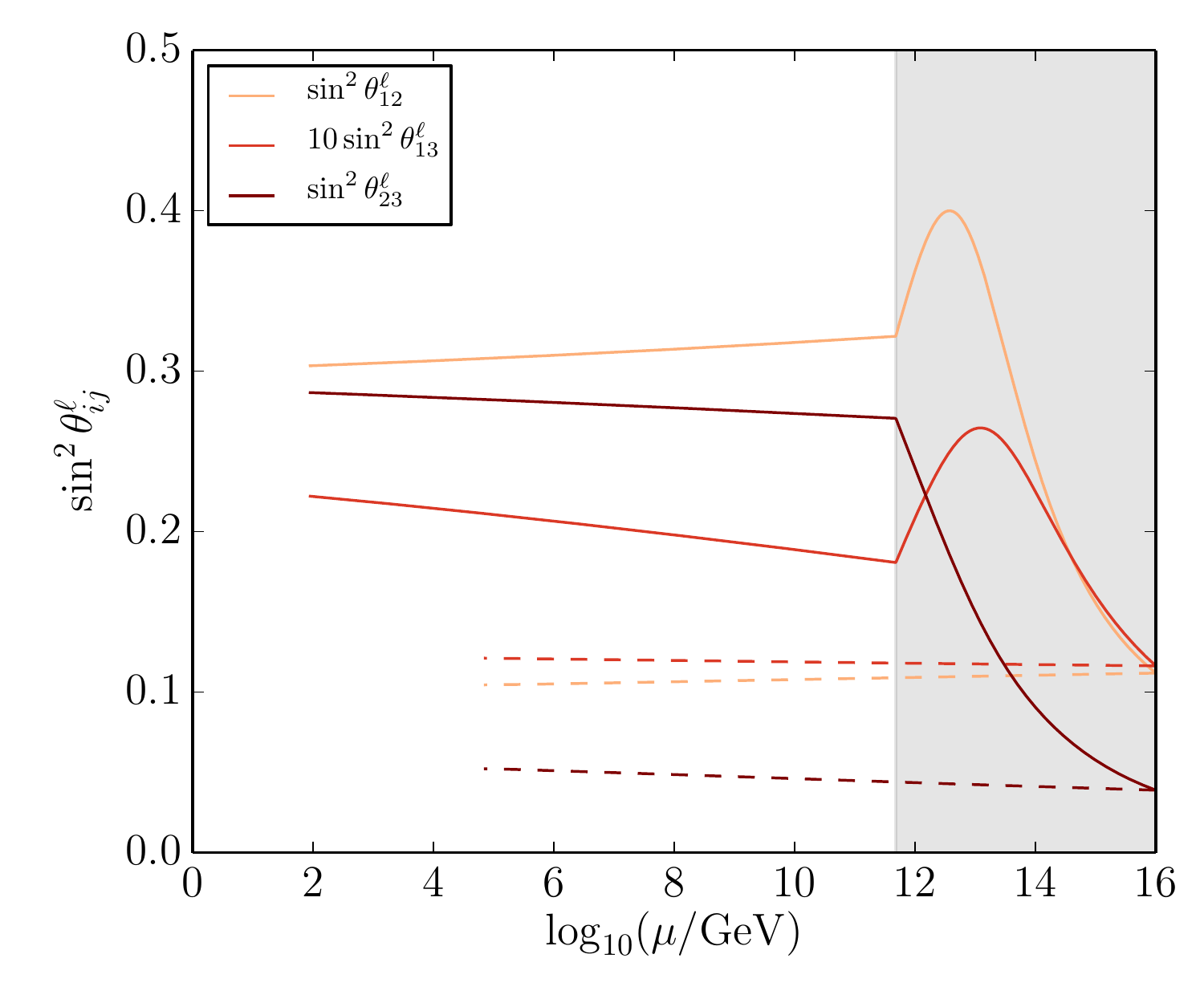}&

\includegraphics[width=0.49\textwidth]{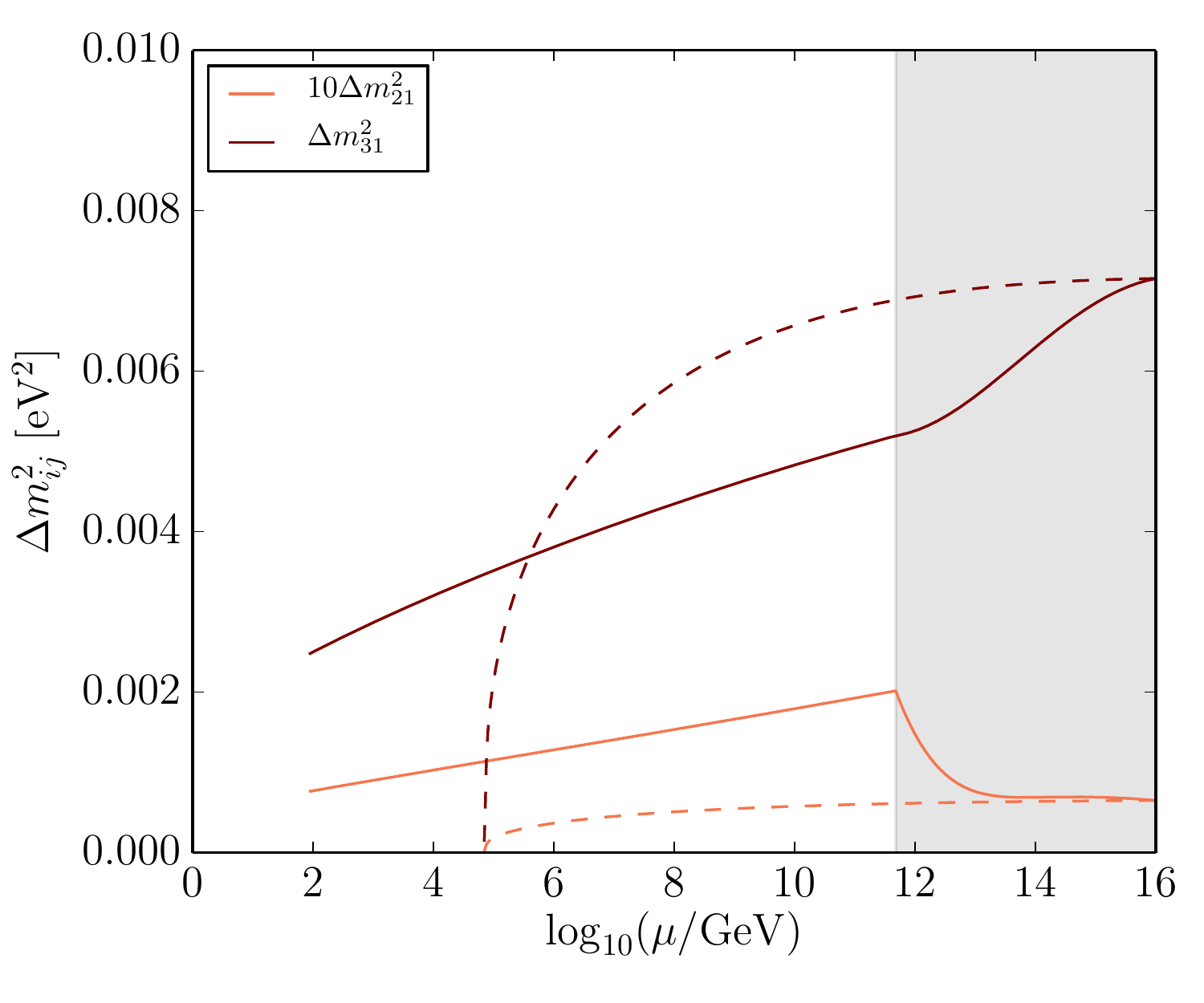}
\end{tabular}

\caption{\label{fig:running}RG running of the quark masses (upper left panel), charged-lepton masses (upper right panel), leptonic mixing angles (lower left panel), and neutrino mass-squared differences (lower right panel) in the extended model with normal neutrino mass ordering. The solid (dashed) curve shows the RG running with (without) the intermediate gauge group between $M_\mathrm{GUT}$ and $M_\mathrm{I}$, denoted by the shaded area.}
\end{figure*}

In figure~\ref{fig:running}, the RG running of the fermion observables (except the quark mixing parameters since they exhibit small RG running) from $M_\mathrm{GUT}$ down to $M_\mathrm{Z}$ for the best-fit parameter values of extended model with NO are presented, with the dashed curves showing the RG running without the intermediate gauge group. That is, assuming the SM-like model discussed in section~\ref{2HDM} for the whole energy range and with the same best-fit parameter values at $M_\mathrm{GUT}$. Note that since particle mass states are not well defined before electroweak symmetry breaking, the parameters above $M_\mathrm{Z}$ are to be considered as effective parameters of the model. They are computed by applying the matching conditions to the parameters of the PS model in order to transform them to the parameters of the 2HDM from which the observables can be calculated. It is evident that the intermediate gauge group has a significant effect on the RG running. The quark masses display a diverging trend as the energy scale decreases from $M_\mathrm{GUT}$ in the case of no intermediate gauge group. In fact, the parameters diverge so that the system of equations has no solution below a certain energy (which is why the dashed curves do not cover the full energy range even for the other observables).  For the charged-lepton masses, the leptonic mixing angles, and the neutrino mass-squared differences, the difference in slope between the PS and 2HDM models is more pronounced. Particularly, the leptonic mixing angles exhibit large RG running in the PS model, but almost no RG running in the 2HDM model.

As a consistency check, one can observe that the RG running below $M_\mathrm{I}$ in the case of an intermediate gauge group is of a similar form as that without an intermediate gauge group (since it is the same equations that are being solved in these two cases, with different initial conditions). With the intermediate gauge group, the diverging effect is offset to lower energies so that the divergence does not occur within the energy region under study, since the RGEs of the 2HDM are applied over a smaller energy range than if there were no intermediate gauge group. In fact, if one extends the RG running to energies well below $M_\mathrm{Z}$, the diverging behavior occurs at a lower energy scale. 

In order to describe the general behaviour of the RG running, particularly the quark and charged-lepton masses since these are linear in the Yukawa couplings, one can approximate the RGEs listed in ref.~\cite{Meloni:2016rnt}, by their leading terms. For the energy region between $M_\mathrm{GUT}$ and $M_\mathrm{I}$, the leading term is in all cases the one involving the gauge couplings. To an accuracy of a few percent, the RGEs can be approximated by 
\begin{equation}
16\pi^2\frac{\mathrm{d}Y_i}{\mathrm{d}t}\simeq f_i(g_j(t))Y_i(t),
\end{equation}
where $t=\mathrm{ln}(\mu/\mathrm{GeV})$, the subscript $i$ denotes the Yukawa coupling matrix in question and $f_i$ denotes a function of the three gauge couplings $g_j$. The same applies to the RGEs in the 2HDM model for $Y_u$ and $Y_d$, but for $Y_e$ the leading term is the one involving $Y_d$ such that the RGE may be approximated by
\begin{equation}
16\pi^2\frac{\mathrm{d}Y_e}{\mathrm{d}t}\simeq 3\mathrm{tr}(Y_d(t)^\dagger Y_d(t))Y_e(t).
\end{equation}
In the energy region below $M_\mathrm{I}$, the approximation is not quite as good with the error of the charged-fermion masses at $M_\mathrm{Z}$ between $10~\%$ and $30~\%$. The fact that the RGEs for the quark and charged-lepton Yukawa matrices have different leading terms explains why the masses exhibit such different RG running. The mixing parameters and neutrino mass-squared differences cannot be well approximated by the leading terms, since these observables are related to the Yukawa couplings in a more complicated and non-linear way.

The other two works that take into account the effect of the intermediate gauge group, refs.~\cite{Meloni:2014rga,Meloni:2016rnt}, agree with our conclusion that it has a significant effect on the RG running of the parameters and thus on the fit itself. However, we find a considerably different behavior of the RG running of the parameters as well as increased difficulty in fitting the $\mathrm{SO}(10)$ models to the data. Comparing the effect of the intermediate gauge group on the RG running in our work with that of ref.~\cite{Meloni:2016rnt}, we find a considerably closer similarity between the RG running behaviour of the parameters below $M_\mathrm{I}$ with that in the absence of an intermediate gauge group. In comparison to ref.~\cite{Joshipura:2011nn}, they, like us, concluded that the easiest model to fit to (out of the ones considered above) is the extended model with NO, in agreement also with ref.~\cite{Dueck:2013gca}. The latter work also found that IO is more difficult to fit to than NO. However, they could find considerably better fits than we have found in all cases, since they do not take into account the intermediate gauge group, which increases the complexity of the problem considerably and complicates the fit. Of course, it must be noted that we cannot ensure that we have found the global minimum and cannot with complete certainty rule out the other three cases for which no acceptable fit was found. Furthermore, there may be other effects which may act to improve the ability to fit the models to the measured values of the observables, such as higher-order terms in the RGEs and threshold corrections to the RG running~\cite{Babu:2015bna}. Other corrections may also arise from variations in some of the assumptions regarding the SM-like model. Such corrections may be particularly interesting for the minimal model with normal neutrino mass hierarchy, since this case is not too far from having a reasonable fit. One possibility may be to include the effects of type-II seesaw, particularly since the most tension was found in the leptonic mixing sector.

\section{Summary and conclusions}\label{conclusion}
We have performed numerical fits to two different non-SUSY $\mathrm{SO}(10)$ models, namely the minimal model and the extended model, which differ by the inclusion of a $\mathbf{120}_H$ representation in the Yukawa sector of the Lagrangian. The fits were performed with both NO and IO, and assuming a type-I seesaw mechanism for neutrino mass generation. The results of the fits show that out of the four cases considered, only the extended model with NO is viable with $\chi^2\simeq 18.6$. One reason for the difficulty in finding acceptable fits of the models is the extra complexity introduced by the intermediate gauge group, which has been shown to have a considerable effect on the RG running as can be seen from the change in slope at $M_\mathrm{I}$ in figure~\ref{fig:running}. Another reason for the difficulty in the fitting procedure is the fact that the best-known value of $\theta^\ell_{23}$ is now in the higher octant, whereas a value in the lower octant (as previously predicted) would considerably improve the fit. However, before definitely ruling out the three cases that were unable to accommodate the measured values of the observables at $M_\mathrm{Z}$, one should investigate the effects of higher-order terms as well as threshold corrections.

\section*{Acknowledgments}

We would like to thank Sofiane Boucenna, Davide Meloni, Stella Riad, and Shun Zhou for useful discussions. T.O.~acknowledges support by the Swedish Research Council (Vetenskapsr\r{a}det) through contract No.~2017-03934 and the KTH Royal Institute of Technology for a sabbatical period at the University of Iceland. M.P.~thanks ``Stiftelsen Olle Engkvist Byggm{\"a}stare'' and ``Roland Gustafssons Stiftelse f{\"o}r teoretisk fysik'' for financial support. Numerical computations were performed on resources provided by the Swedish National Infrastructure for Computing (SNIC) at PDC Center for High Performance Computing (PDC-HPC) at KTH Royal Institute of Technology in Stockholm, Sweden under project number PDC-2017-115. 


%

\end{document}